\documentclass[12pt]{article}

\usepackage{amscd}

\usepackage{verbatim}

\usepackage{amssymb}

\usepackage{amsmath}

\newcommand{\be}{\begin{equation}}
\newcommand{\ee}{\end{equation}}
\newcommand{\ben}{\begin{eqnarray}}
\newcommand{\een}{\end{eqnarray}}
\newcommand{\ba}{\begin{eqnarray}}
\newcommand{\ea}{\end{eqnarray}}

\newcommand{\bi}{\begin{itemize}}
\newcommand{\ei}{\end{itemize}}

\begin{document}

\begin{center}

\vspace{24pt} { \large \bf Stability analysis of (a class of) anisotropic spacetimes } \\

\vspace{30pt}

\vspace{30pt}

{\bf Bhavesh Khamesra\footnote{bhaveshkhamesra@students.iiserpune.ac.in}}, {\bf V
Suneeta\footnote{suneeta@iiserpune.ac.in}}

\vspace{24pt} 
{\em  The Indian Institute of Science Education and Research (IISER),\\
Pune, India - 411008.}

\end{center}

\date{\today}
\bigskip

\begin{center}
{\bf Abstract}\\
\end{center}
\vskip 0.5cm
We consider spherically symmetric spacetimes sourced by a fluid with pressure anisotropy in the radial direction. We use gauge-invariant perturbation theory to study the stability of this class of spacetimes under axial perturbations. We apply our results to three diverse examples. Two examples arise as endpoints of collapse of a ball of fluid --- one describes a well-behaved stellar interior and the other has a naked singularity. We prove the stability of the stellar interior both with respect to Dirichlet and quasinormal mode boundary conditions on the perturbation. Surprisingly, the naked singularity is also stable under axial perturbations. Lastly, we take the example of anisotropic cosmology to show that in this case, the relevant perturbations are those in which the direction of anisotropy is also perturbed.

\newpage

\section{Introduction}

Spacetimes sourced by anisotropic matter sources appear in diverse contexts in gravitational physics. One common source is a fluid with anisotropy in pressure. We give a few examples below of such spacetimes (for more detailed examples/references, see the comprehensive review by Herrera and Santos \cite{herrera}).: \\

a) \textbf{Stellar interiors:} There are examples of spherically symmetric solutions to Einstein's equations with an anisotropic fluid source with pressure anisotropy in the radial direction (as compared to the angular directions). We call these spacetimes anisotropic due to anisotropy of pressure (and consequent anisotropy of metric components in radial and angular directions) --- this terminology differs from that in cosmology where isotropy generally implies spherical symmetry. There are nonsingular spacetimes in this class which have been used as a model for realistic stellar interiors \cite{florides}, \cite{gleiser}. Toy models of collapsing matter sourced by such fluids have also been studied --- the collapse can lead to spacetimes with a naked singularity \cite{joshinarayan}. Such toy models can be used to study various pathologies of naked singularities. Spacetimes sourced by rotating anisotropic fluids have been used to model the interior of a rotating star such that the metric matches to the Kerr metric outside the star \cite{kyriakopoulos}. Although the spacetimes in \cite{kyriakopoulos} have ring singularities, this approach is interesting in view of the fact that to date, there is no example of a stellar interior sourced by a perfect fluid, that matches (with standard matching conditions) to the Kerr metric outside the star. There are other related compact objects that have been proposed with anisotropic interiors, such as gravastars \cite{mazur}. Recently, it has also been found that equations of many-body astrophysical systems with spherical symmetry studied in the Post-Newtonian approach resemble those of the above-mentioned anisotropic fluid sources \cite{pedraza}. \\

b) \textbf{Cosmology}: Spacetimes sourced by anisotropic fluids also find application in cosmology. In a situation in which it is not possible to find one comoving frame in which different cosmological fluids are all at rest, it is possible to go to a frame where the energy-momentum tensor of the multi-fluid system can be rewritten as the energy-momentum tensor of a single anisotropic fluid \cite{Letelier, Letelier1}. Specific examples of such cosmological spacetimes, with applications to the study of voids are given in \cite{bayin}. A more recent application of this framework has been to attempt to explain the observed cosmological anisotropy from Planck \cite{harko}.\\

c)\textbf{ Gauge/Gravity duality} : From recent experimental results from the relativistic heavy ion collider (RHIC) on the quark-gluon plasma, it is known that the plasma is initially anisotropic (with a different pressure in the beam direction as compared to the transverse direction) but quickly becomes isotropic. The Gauge/gravity duality relates this plasma to a suitable gravitational spacetime sourced by anisotropic matter such that string theory on the spacetime can be used to glean information about the plasma (for a review, see \cite{solana}). In this context, the matter source for the spacetime ranges from an anisotropic fluid \cite{janik} to axions (which are responsible for the anisotropy) and dilatons \cite{mateos}.
\vskip 0.5cm

In all these diverse settings in which anisotropic spacetimes appear, there is one question of \emph{universal} importance. This is the question of stability of the spacetime to small perturbations (of the metric and the matter). We would also like to know if anisotropy grows/decreases due to specific perturbations. For spacetimes describing stellar interiors, we require the spacetime to be stable to make physical sense. For the spacetime dual to the quark gluon plasma, we need an instability that causes the spacetime to `isotropize'. For specific classes of perturbations of anisotropic stellar interiors, a stability analysis has been done in \cite{ruggeri} (assuming a specific equation of state), and \cite{chan2} and \cite{chan3} (no specific equation of state assumed). These studies bring out the connection between the initial pressure anisotropy parameter and the stability issue. The astrophysical relevance of this specific class of perturbations and these results is discussed in \cite{chan1}. A partial stability study of some anisotropic spacetimes (under radial perturbations) in the context of stellar interiors has also been done in \cite{gleiser2}. It has been found that anisotropic stars are more stable than isotropic ones if the tangential pressure is greater than the radial pressure. A nice explanation for this can be found in \cite{herreragrg}, \cite{herrerapla} --- in \cite{herreragrg}, a local version of the Tolman-Whittaker mass \cite{tolman}, \cite{whittaker} is defined. Further, it is shown in \cite{herrerapla} using this local mass function, that for a collapsing (anisotropic) fluid sphere, if the radial pressure is greater than the tangential pressure, collapse is accelerated (signifying instability). In the context of non-rotating gravastars, analysis reveals stability under axial perturbations \cite{chirenti}, however rotating gravastars can have instabilities \cite{cardoso}, \cite{chirenti2}. In cosmology, stability of cosmologies sourced by viscous fluids (Bianchi I spacetimes) have been extensively discussed, c.f. \cite{perko, den, miedema}. In the context of gauge-gravity duality, thermodynamic instabilities of the anisotropic spacetime in \cite{mateos} have been discussed in \cite{mateos2} (see also \cite{cheng}, \cite{cheng1}).

\vskip 0.5cm

It would be desirable to do a systematic stability analysis for different classes of anisotropic spacetimes. In this work, we will deal with spherically symmetric spacetimes which are sourced by fluids with pressure anisotropy in the radial direction. This class of spacetimes allows for a decomposition of perturbations into standard vector and scalar perturbations (i.e., given in terms of scalar/vector spherical harmonics on the two-sphere). Due to decoupling of the Einstein equations for these two classes of perturbations, we can consider each case separately. We will discuss the vector or `axial' case in this paper.

In the next section, we review basic results about anisotropic fluids and the Einstein equations for such sources. In section III, we discuss axial perturbations of anisotropic spacetimes using a formulation of Ishibashi and Kodama \cite{ishikodama}
which uses gauge-invariant perturbation variables. In section IV---VI, we discuss applications of this formalism to the study of some of the diverse settings mentioned in the introduction --- stellar interiors, pathological spacetimes with naked singularities and cosmology. Finally, we conclude with a summary of our results and interesting future directions.

\section{Anisotropic fluids}
In this section, we will consider four dimensional spacetimes sourced by fluids with a pressure anisotropy in the radial direction. The tangential (angular) pressures are assumed to be the same, and therefore such spacetimes will be spherically symmetric. We can write the spacetime metric as
\begin{eqnarray}
ds^{2} = -e^{2\nu}dt^{2}+e^{2\psi}dr^{2}+\tilde R^{2}d\Omega^{2},
\label{2.1}
\end{eqnarray}
where $d\Omega^{2}$ is the standard metric on a two-sphere of unit radius. $\nu$, $\psi$ and $\tilde R$ are functions of only $r$ and $t$ due to spherical symmetry.
In these coordinates, we can write the energy momentum tensor of the fluid source as
\begin{eqnarray}
T_{\nu}^{\mu}=\left(\begin{array}{cccc}
-\rho & 0 & 0 & 0\\
0 & p_{r} & 0 & 0\\
0 & 0 & p_{t} & 0\\
0 & 0 & 0 & p_{t}
\end{array}\right)
\label{2.1a}
\end{eqnarray}
where \(\rho\) is the energy density of the fluid, \( p_{r}\) is the radial(normal) pressure and \(p_{t}\) is the tangential (angular) pressure. These are all functions of the radial variable $r$ and time $t$. We also need to specify two equations of state for $p_r $ and $p_t $ as functions of $\rho$.

This energy momentum tensor can be written in arbitrary coordinates as
\begin{eqnarray}
 T_{\nu}^{\mu}= (\rho + p_t)u_{\nu}u^{\mu}+(p_{r} - p_{t})s_{\nu}s^{\mu} + p_{t}\delta_{\nu}^{\mu} .
 \label{2.2}
\end{eqnarray}\\

In coordinates in which the fluid is at rest, the fluid four-velocity $ u^{\mu}=e^{-\nu}\delta^{\mu}_{0}$  and $s^{\mu}=e^{-\psi}\delta^{\mu}_{0}$ is a unit vector in the radial direction giving the direction of anisotropy. These quantities satisfy the relations $u^{\mu}u_{\mu}=-1$, $s_\mu s^\mu =1$ and $s_\mu u^\mu=0 $. It can be seen that (\ref{2.2}) then reduces to (\ref{2.1a}).

Due to spherical symmetry, $\rho$, $p_r$ and $p_t$ are functions only of $r$ and $t$. In what follows, we will denote partial derivatives with respect to $r$ by primes, and derivatives with respect to $t$ by dots.
The energy-momentum tensor conservation equations are (in units $c,~~ G = 1$):
\begin{equation}
 \dot{\rho}+(p_{r}+\rho)\dot{\psi}+(p_{t}+\rho)\frac{2\dot{\tilde R}}{\tilde R}=0.
 \label{2.2a}
\end{equation}
\begin{equation}
 p'_{r}+(p_{r}+\rho)(\nu)'+(p_r-p_t)\frac{2\tilde R'}{\tilde R}  =0.
 \label{2.2b}
\end{equation}
Given equations of state for $p_r $ and $p_t $ in terms of $\rho$, we can obtain an equation for $\rho$ and the function $\tilde R$.
The Einstein equations are
\begin{equation}
\frac{2m'}{\tilde R^2 \tilde R'}=8 \pi\rho .
\label{2.3}
\end{equation}
\begin{equation}
 \frac{2\dot{m}}{\tilde R^2 \dot{\tilde R}} = - 8 \pi p_r .
 \label{2.4}
\end{equation}
\begin{equation}
(\nu)' \dot{\tilde R} + \tilde R' \dot \psi = \partial_t \tilde R' .
\label{2.4a}
\end{equation}
\begin{eqnarray}
&&\left((\nu)''e^{-2\psi}-\ddot{\psi}e^{-2\nu}-\frac{\ddot{\tilde R}}{\tilde R}e^{-2\nu}+\frac{\tilde R''}{\tilde R}e^{-2\psi}\right)\tilde R^2+ \left([(\nu)']^{2}e^{-2\psi}-\dot{\psi}^2e^{-2\nu}\right)\tilde R^2+ \nonumber \\
&&\left(\dot{\nu}\dot{\psi}e^{-2\nu}-(\nu)'\psi'e^{-2\psi}\right)\tilde R^2 +\left((\nu)'\tilde R'e^{-2\psi}+\dot{\nu}\dot{\tilde R}e^{-2\nu}- \dot{\psi}\dot{\tilde R}e^{-2\nu}-\psi'\tilde R'e^{-2\psi}\right)\tilde R \nonumber \\&&~~~~~~~~~~~~~~~~~~~~~~~~~~~~~~= 8 \pi p_t \tilde R^2 .
\label{2.5}
\end{eqnarray}

where $m(r,t)$ is the Misner Sharp mass function \cite{misner}
\begin{equation} m(r,t) =\frac{\tilde R}{2}(1+\dot{\tilde R}^2 e^{-2\nu}-\tilde R'^2 e^{-2\psi}).
\label{2.6}
\end{equation}
The Misner-Sharp mass function, roughly speaking, gives the mass enclosed inside a spherical ball of radius $\tilde R$. Specifically, if the interior of a spherical star is modelled by a (static) fluid ball with radial pressure vanishing on the fluid boundary, then $m$ yields the Schwarzschild mass when evaluated on the boundary surface. In this case, we can also match the interior spacetime to a Schwarzschild spacetime in the exterior. The equations in the static case are:
 \\
\begin{equation}
1 + 2\psi'\tilde R'\tilde R e^{-2\psi}- \tilde R'^{2}e^{-2\psi} - 2 \tilde R {\tilde R}'' e^{-2\psi} = 8 \pi \rho \tilde R^{2}. \label{2.7}
\end{equation}
\begin{equation}
 1 - 2(\nu)'\tilde R \tilde R'e^{-2\psi}  - \tilde R'^2e^{-2\psi}= - 8 \pi p_r \tilde R^2 .
 \label{2.8}
\end{equation}
\begin{equation}
 (\nu)''+\frac{\tilde R''}{\tilde R}+[(\nu)']^2-(\nu)'\psi'+
 (\nu)'\frac{\tilde R'}{\tilde R}-\psi'\frac{\tilde R'}{\tilde R} = 8 \pi p_t e^{2\psi} .
 \label{2.9}
\end{equation}

The above equations and the conservation equation (\ref{2.2b}) can be solved to obtain the functions $\nu,\psi$ ,$\tilde R$ and $\rho$ (in the static case, (\ref{2.2a}) is trivially satisfied).

We will discuss specific solutions to the Einstein equation in subsequent sections. However, before this, we would like to discuss some standard results concerning perturbations of solutions.

\section{Perturbation theory and gauge invariant variables of Ishibashi/Kodama}
As we discussed in the introduction, anisotropic spacetimes occur in physics in various contexts, however, the question of stability of the spacetime to perturbations is of common interest in these diverse situations. Perturbation theory for the spherically symmetric backgrounds (\ref{2.1}) can be efficiently organized by classifying metric perturbations based on their tensorial behaviour on the two-sphere. We will summarize some results from the perturbation formalism of Ishibashi and Kodama (IK) \cite{ishikodama}, which will be used in this paper. The notation is as follows:
Consider the four dimensional spacetime  (\ref{2.1}) written as
\begin{equation}
g_{\mu\nu}dx^{\mu}dx^{\nu} = g_{ab}(y)dy^{a}dy^{b}+\tilde{R}^2 (y)\gamma_{ij}(z)dz^{i}dz^{j} .
\label{3.1}
\end{equation}

$g_{ab}(y)$ is the Lorentzian metric defined on the $r-t$ submanifold
and $\gamma_{ij}(z)$ is the standard metric on the unit sphere $S^2$. Henceforth, we shall assume that (\ref{3.1}) solves Einstein equations with source given by (\ref{2.1a}).

We adopt the following convention to differentiate tensors associated with these submanifolds:\\
Greek indices denote tensor indices on the four dimensional spacetime;
Latin indices $a, b$ denote tensor indices on the $r-t$ submanifold and
Latin indices $i, j$ denote tensor indices on $S^2$. \\
 \indent We denote covariant derivatives, connection coefficients and Riemann curvature tensors for the full spacetime, the $r-t$ submanifold and the sphere with metric
$\gamma_{ij}$ as\\
\[g_{\mu\nu}dx^{\mu}dx^{\nu} \to \nabla_{\mu}, \Gamma^{\alpha}_{\beta\gamma}, R_{\mu\nu\alpha\beta}\]
\[g_{ab}dy^{a}dy^{b} \to D_{a}, \bar{\Gamma}^{a}_{bc}, \bar{R}_{abcd}\]
\[\gamma_{ij}dz^{i}dz^{j} \to \hat{D}_{i}, \hat{\Gamma}^{i}_{jk},\hat{R}_{ijkl}\]\\

$\hat{R}_{ij}=K\gamma_{ij}$, where the constant $K=1$ corresponds to the sectional curvature of $S^2$.

Let us now consider a perturbation of the background spacetime as $\delta g_{\mu\nu}=h_{\mu\nu}$. Then, in our notation, $h_{ab}$ is a perturbation with no tensor indices on the sphere. Thus, it is a scalar with respect to coordinate transformations on the sphere, and such a perturbation can be decomposed into scalar spherical harmonics. Similarly, from standard perturbation theory, $h_{ai}$ (vector on the sphere) and $h_{ij}$ (two-tensor on the sphere) can be written as \cite{ishikodama}:
\begin{eqnarray}
&& h_{ai}=\hat{D}_i h_a +h^{(1)}_{ai}. \nonumber \\
&& h_{ij}=h_{T\;ij}^{(2)}+2\hat{D}_{(i}h^{(1)}_{T\;j)}+h_L\gamma_{ij} + \hat{L}_{ij}h^{(0)}_T .
\label{3.2}
\end{eqnarray}\\
\noindent where $\hat{D}^j h^{(2)}_{T\;ij}= h^{(2)i}_{T\quad i}=0,\, \hat{D}^{i}h^{(1)}_{T\; i}=0 \,\, $and$ \,\, \hat{D}^i h^{(1)}_{ai}=0 $. A factor of $2$ comes in the second term as the symmetrization of indices bracket has a factor of $\frac{1}{2}$.
A similar decomposition can be done for the perturbed energy-momentum tensor. $\delta T_{ab}$ can be written in terms of the scalar spherical harmonics. For the other perturbations,
\begin{eqnarray}
&& \delta T_{ai}=\hat{D}_i \delta T_a +\delta T^{(1)}_{ai} .\nonumber \\
&& \delta T_{ij}=\delta T_{T\;ij}^{(2)}+2\hat{D}_{(i}\delta T^{(1)}_{T\;j)}+\delta T_L\gamma_{ij} + \hat{L}_{ij}\delta T_T .
\label{3.3}
\end{eqnarray}\\
\noindent
Here $\hat{D}^j \delta T^{(2)}_{T\;ij}= \delta T^{(2)i}_{T\quad i}=0,\, \hat{D}^{i}\delta T^{(1)}_{T\; i}=0 \,\, $and$ \,\,\hat{D}^i \delta T^{(1)}_{ai}=0 $.
The perturbation variables $h_{a}, h_{L}, h^{(0)}_T ,\delta T_a ,\delta T_L , \delta T_T  $ can be expanded again
in terms of scalar spherical harmonics, $h^{(1)}_{ai}, h^{(1)}_{T\; i}, \delta T^{(1)}_{T\;j}, \delta T^{(1)}_{ai} $ can be expanded in terms of the divergence-free (transverse) vector spherical harmonics, and $h^{(2)}_{T\;ij}, \delta T_{T\;ij}^{(2)}$ can be expanded in terms of the transverse traceless (TT) tensor spherical harmonics. However there are no TT tensor harmonics on the two-sphere, so we can set the latter pieces to zero.

The linearized Einstein equations for the perturbed spacetime decouple for the perturbations written in terms of the vector spherical harmonics (`axial' perturbations), and the scalar spherical harmonics (`polar' perturbations). The full stability problem is dealt with by considering these two classes of perturbations separately. In this paper, we deal with the axial perturbations. We therefore set $h_{ab}$ and all perturbation variables given in terms of the scalar spherical harmonics to zero. The perturbations of the metric and the fluid are not gauge-invariant (under coordinate transformations of the spacetime). We will use the  gauge-invariant combinations of these variables constructed by IK:
\begin{eqnarray}
F^{(1)}_{ai} &:= &h^{(1)}_{ai}-\tilde R^2 D_{a} \left(\frac{h^{(1)}_{Ti}}{\tilde R^2}\right) . \nonumber \\
\tau^{(1)}_{ai}& :=&\delta T^{(1)}_{ai}- p_{t} h^{(1)}_{ai} \nonumber .\\
\tau^{(1)}_{ij}&:= &2\hat{D}_{(i} \delta T^{(1)}_{Tj)}-2p_{t} \hat{D}_{(i} h^{(1)}_{Tj)} .
\label{3.4}
\end{eqnarray}
The divergence-free vector spherical harmonics are defined by the equation
$$( \gamma^{kl} \hat D_{k} \hat D_{l} + k_{v}^{2} ) \mathcal{V}_{i} = 0,$$
where $k_{v}^{2} = l(l+1) - 1$, $l=1,2,....$. Let $m_V := k^{2}_{v}-1=l(l+1)-2$. For $m_V > 0$, we can write the gauge-invariant variables (\ref{3.4}) in terms of the vector spherical harmonics as
\begin{eqnarray}
F^{(1)}_{ai}&=&\tilde R F_a\mathcal{V}_i, \quad \tau^{(1)}_{ai}=\tilde R \tau_a\mathcal{V}_i, \nonumber \\ \tau^{(1)}_{ij}&=& \tilde R^2 \tau_T\mathcal{V}_{ij}.
\label{3.5}
\end{eqnarray}
Here $\mathcal{V}_{ij} = - (1/k_v )\hat D_{(i}\mathcal{V}_{j)}$.
The perturbed Einstein equations in terms of these variables (for $m_V > 0$) are:
\begin{eqnarray}
\frac{1}{\tilde R^{3}}D^b\left\{\tilde R^{4}\left[D_b\left(\frac{F_a}{\tilde R}\right)-D_a\left(\frac{F_b}{\tilde R}\right)\right]\right \}-\frac{m_V}{\tilde R^2}F_a&=&-16\pi \tau_a ; \nonumber \\
 \frac{k_V}{\tilde R^2}D_a(\tilde R F^a)&=& -8\pi \tau_T .
 \label{3.6}
\end{eqnarray}
As has been discussed in \cite{ishikodama}, the $m_V = 0$ case can be analyzed separately and does not correspond to dynamical perturbations. Therefore, we will not consider it. Note that for the axial perturbations, the trace $Tr(h) = g^{ai} h_{ai} + g^{ij} h_{ij} = 0$, as can be seen from the form of (\ref{3.2}) for this class of perturbations, and the fact that for the background, $g^{ai} = 0$.

We also have the perturbed energy-momentum tensor conservation equation
\begin{equation}
\delta(\nabla_{\mu}\ T^{\mu\nu})=\nabla_{\mu}\delta T^{\mu \nu} + \delta\Gamma^{\mu}_{\mu\alpha}T^{\alpha\nu}+\delta\Gamma^{\nu}_{\mu\alpha}T^{\alpha\mu} = 0.
\label{3.7}
\end{equation}
The energy-momentum tensor of the background is given by (\ref{2.1a}, \ref{2.2}). Considering vector perturbations, we see that (\ref{3.7}) is trivially satisfied for $\nu = a$. For $\nu = i$, in terms of gauge-invariant variables, the equation can be manipulated into the very simple form
\begin{equation}
 D_{a}(\tilde R^{3}\tau^{a})+\frac{m_V}{2k_{V}}\tilde R^{2} \tau_{T} = 0.
 \label{3.8}
\end{equation}
For the anisotropic fluid (\ref{2.2}),
\begin{equation}
\delta T_{\mu\nu} = 2(\rho+p_{t})u_{(\mu}\delta u_{\nu)} + 2(p_{r} - p_{t})s_{(\mu}\delta s_{\nu)}+p_t h_{\mu\nu}.
\label{3.9}
\end{equation}

Let $\delta u_{i}=\tilde{\alpha}\mathcal{V}_{i}$ and $\delta s_{i}=\tilde{\beta}\mathcal{V}_{i} $. $\delta u_{a}=\delta s_{a} =0 $ because these will
behave like scalars on \(S^{2}\). Similarly, pressure and density perturbations are scalars and we therefore do not consider them. We have considered the most general vector perturbations of the energy-momentum tensor. The perturbation $\delta s_{i}$ corresponds to perturbing the `direction of anisotropy'. Whether such a perturbation is to be considered depends on the physical system. In later sections, we will consider three examples, two where this can be set to zero, and one where this perturbation may be physically relevant. A third possibility is to identify the term proportional to $\delta s_{i}$  with a standard shear term which is of the form $$\sigma^{\alpha\beta}=\frac{1}{2}\nabla_\mu u^{\alpha}(u^{\mu}u^{\beta}+g^{\mu\beta})+ \frac{1}{2} \nabla_\mu u^{\beta}(u^{\mu}u^{\alpha}+g^{\mu\alpha})-\frac{1}{3}\theta (u^{\alpha}u^{\beta}+g^{\alpha\beta});$$ where one of the indices $\alpha, \beta$ is $r$ and the other $i$ (an angular coordinate index), and $\theta = \nabla_{\mu}u^{\mu}$. Such shear terms lead to dissipation. We will not consider such a possibility in this paper.

From (\ref{3.4}), (\ref{3.5}) and (\ref{3.9}), it is easy to see that $\tau_T = 0$. $\tilde R \tau_{0} = - (\rho + p_t ) e^{\nu} \tilde  \alpha $ and $ \tilde R \tau_{1} = (p_r - p_t ) e^{\psi}  \tilde \beta$; where the subscript $0$ refers to the $t$ coordinate and $1$ refers to the $r$ coordinate. Then (\ref{3.8}) becomes
\begin{eqnarray}
&&\tilde R^3 [\dot \tau^{0} + \tau^{1'} ] + 3 \tilde R^{2} \dot{ \tilde R }\tau^{0} +  \dot \nu \tilde R^3 \tau^{0} +  \dot \psi \tilde R^{3} \tau^{0} +
\nu' \tilde R^3 \tau^{1} + \nonumber \\ &&  \psi' \tilde R^{3} \tau^{1} + 3 \tilde R^{2} \tilde R' \tau^{1} = 0.
\label{3.10}
\end{eqnarray}

For the special case of a static background metric, the perturbed conservation law becomes
\begin{eqnarray}
 (p_t+\rho)\dot{\tilde \alpha } e^{\psi-\nu} +(p_r-p_t)\tilde \beta'+\left(p_r'-p_t'\right)\tilde \beta +(p_r-p_t)\tilde \beta\left(\nu'+\frac{2R'}{R}\right)=0.~~~~~~
\label{3.10a}
\end{eqnarray}

We notice here that if the direction of anisotropy is not perturbed (i.e., $\tilde \beta = 0$), then this equation implies that $\dot{ \tilde \alpha} = 0$ as well. Thus the matter perturbation has no time dependence and cannot excite time-dependent metric perturbations.  In a standard `modal analysis' of metric/matter perturbations, i.e., if we let $\tilde \alpha = e^{\lambda t} \alpha$, then we get that in this case $\alpha = 0$.

As shown by IK, the perturbed Einstein equations and the perturbed energy-momentum tensor equation can be combined to yield a single equation for a function, the `master equation'.
From the second equation in (\ref{3.6}) and (\ref{3.8}), we first obtain $D_a(\tilde R F^a-\frac{16\pi}{m_V}\tilde R^3\tau^a)=0 $.
Hence we can write
\begin{equation}
\epsilon^{ab}D_{b} \tilde \Omega = \tilde R F^a-\frac{16\pi}{m_V}\tilde R^3 \tau^a .
\label{3.11}
\end{equation}
Substituting for $F^a $ from (\ref{3.11}) into the first equation in (\ref{3.6}) yields (after manipulations) the `master equation'
\begin{eqnarray}
 \tilde{R}^2 D_a\left(\frac{1}{\tilde{ R}^2}D^a\tilde{\Omega}\right)-\frac{m_V}{\tilde{R}^2}\tilde{\Omega}=-\frac{16\pi}{m_V}\tilde{R}^2\epsilon^{ab}D_a(\tilde R \tau_b).
 \label{3.12}
\end{eqnarray}
We end this section by noting again that if the background is static and $\tilde \beta=0$, this implies that in a standard modal stability analysis, the right-hand-side of (\ref{3.12}) vanishes. In this situation axial perturbations of the spacetime cannot excite (or be excited by) the fluid perturbations. In spacetimes sourced by perfect fluids (models for stellar interiors) such a decoupling of axial perturbations from fluid modes is well-known \cite{chandraferrari}. We see that the same holds true even in the anisotropic case considered in this paper, provided the background is static and anisotropy direction is constant.

\section{Stellar interiors} We seek to apply the analysis of the previous section to a spacetime sourced by an anisotropic fluid, which could describe a stellar interior. There are an infinity of static, spherically symmetric solutions with the stellar exterior being described by a Schwarzschild spacetime, parametrized by two generating functions (or equivalently, choice of equation of state for $p_r$ and $p_t$) \cite{herreraPRD}. Some specific examples of such stellar interiors are given in \cite{florides}, \cite{gleiser},\cite{bowersliang}. We pick one of these examples to illustrate the axial perturbation stability analysis. The example, due to Florides \cite{florides}, is for a spherically symmetric static stellar interior sourced by a fluid with radial pressure $p_r = 0$. As it is static, the Einstein equations are given in this case by (\ref{2.7})---(\ref{2.9}). In general, for a static interior to be able to match to a Schwarzschild spacetime in the exterior, one only needs the radial pressure to vanish on the stellar surface --- more general examples can be found in \cite{herreraPRD}.

The stellar interior metric of Florides is
\begin{equation}
 ds^2= -\left(1-\frac{2M}{a}\right)e^{\int_a^r \frac{2\mu}{r^2\left(1-\frac{2\mu}{r}\right)}dr}dt^2+\left(1-\frac{2\mu}{r}\right)^{-1}dr^2+r^2 d\Omega^2 ;
 \label{4.1}
\end{equation}
 where $a$ is the radius of the star and $r\leq a$. To ensure the correct signature of metric, we impose the condition $r>2\mu(r)$  for all $r\leq a$. Here
 $$\mu(r)=4\pi\int_0^r \rho(r)r^2dr ; \quad \quad \mu(a)=M;$$
 where $M$ is the total gravitational mass of the sphere. We are free to choose $\rho(r)$ provided some of the inequalities/conditions already mentioned are met. The metric for $r > a$ is just the Schwarzschild metric. We want $ a > 2M$ so that there is no horizon.
The energy momentum tensor is given by (\ref{2.2}) with $p_r = 0$. The tangential pressure $p_t $ is related to the energy density $\rho$ by the `equation of state'
 $$p_t=\frac{\mu\rho}{2r\left(1-\frac{2\mu}{r}\right)}.$$
The choice of equation of state is more for computational simplicity in solving Einstein equations rather than some particular physical consideration. Therefore the resulting spacetime may be thought of as a toy model for studying perturbation theory.

Let us use the results of the previous section to study axial perturbations. The perturbed energy-momentum tensor equation (\ref{3.10}) and the master equation (\ref{3.12}) for the perturbation variable $\tilde \Omega$ are the relevant equations.

The question then is whether perturbations of the direction of anisotropy need to be considered, or whether the anisotropy vector $s^{\mu} $ is constant. Neutron star interiors may have pressure anisotropy in a particular direction \cite{sawyer}--- this is a feature of their composition, and we would not expect gravitational perturbations to excite perturbations of the direction of anisotropy. Thus we set $\tilde \beta = 0$. As the background is static, we can assume the ansatz $\tilde{\Omega}=\Omega e^{i\omega t}$. In such a modal analysis, the master equation (\ref{3.12}) then reduces in the interior $r < a$ to
\begin{eqnarray}
 \left(1-\frac{2\mu}{r}\right)\Omega''+\left(\frac{6\mu}{r^2}-\frac{2}{r}\right)\Omega'+ \left[\omega^2\left(1-\frac{2M}{a}\right)^{-1}e^{-\int_a^r \frac{2\mu}{r^2\left(1-\frac{2\mu}{r}\right)}dr} -\frac{m_V}{r^2}\right]\Omega=0.~~~~~~
 \label{4.2}
\end{eqnarray}
This can be converted to a Regge-Wheeler-type equation \cite{reggewheeler} by introducing a new variable $\Omega=r \Phi$. In terms of $\Phi$, the master equation in the interior becomes ($l \geq 2$):
\begin{eqnarray}
 \left(1-\frac{2\mu}{r}\right)\Phi''+\frac{2\mu}{r^2}\Phi'+\left[\omega^2\left(1-\frac{2M}{a}\right)^{-1}e^{-\int_a^r \frac{2\mu}{r^2\left(1-\frac{2\mu}{r}\right)}dr} -\frac{l(l+1)}{r^2}+\frac{6\mu}{r^3}\right]\Phi=0.~~~~~~
 \label{4.3}
\end{eqnarray}
We introduce a new coordinate for the interior given by
\begin{eqnarray}
d\tilde{r}=\left(1-\frac{2\mu}{r}\right)^{-\frac{1}{2}}\left(1-\frac{2M}{a}\right)^{-\frac{1}{2}}e^{-\int_a^r \frac{\mu}{r^2\left(1-\frac{2\mu}{r}\right)}dr}dr .
\label{4.4}
\end{eqnarray}
The Schrodinger type equation becomes:
\begin{eqnarray} \frac{d^2\Phi}{d\tilde{r}^2}+\left[\omega^2-\left(\frac{l(l+1)}{r^2}-\frac{6\mu}{r^3}\right)\left(1-\frac{2M}{a}\right)e^{\int_a^r \frac{2\mu}{r^2\left(1-\frac{2\mu}{r}\right)}dr}\right]\Phi=0.~~~~~~
 \label{4.5}
\end{eqnarray}

In the exterior $r > a$, we have the Schwarzschild spacetime and the equation in the exterior is the usual Regge-Wheeler equation for the Schwarzschild spacetime. For the exterior, let $d\tilde r = dr/(1-2M/r)$. The interior and exterior equations are both of the form
\begin{equation}
-\frac{d^2\Phi}{d\tilde{r}^2} + V \Phi = \omega^2 \Phi ,
\label{4.6}
\end{equation}
where for $r<a$, $$V = \left(\frac{l(l+1)}{r^2}-\frac{6\mu}{r^3}\right)\left(1-\frac{2M}{a}\right)e^{\int_a^r \frac{2\mu}{r^2\left(1-\frac{2\mu}{r}\right)}dr}.$$
Given the inequalities $r > 2\mu (r)$ and $a> 2M$, and $l \geq 2 $, the potential $V > 0$ for $r < a$.
For $r > a$, $$V = \left(1-\frac{2M}{r}\right) \left(\frac{l(l+1)}{r^2}-\frac{6M}{r^3}\right) > 0$$ is the Regge-Wheeler potential. The potential has a jump discontinuity at $r=a$, a feature also of perfect fluid models of stellar interiors \cite{chandraferrari}.
\\

To address the stability problem, we deal with the following cases:\\

\noindent
\textbf{i)}  $\omega = -i \lambda$ is pure imaginary: \\In this case, the original perturbation variable $\tilde{\Omega}= r \Phi e^{\lambda t}$. Thus, if we have normalizable solutions $\Phi$ for $\lambda > 0$, they would trigger instabilities that grow in time. We multiply (\ref{4.6}) by $\Phi^{*}$ (complex conjugate of $\Phi$) and integrate over the whole spacetime to obtain (after integration by parts),
\begin{eqnarray}
\lim_{d \to \infty} \left(- \Phi^{*} \frac{d\Phi}{d\tilde r}  \right ) \biggr|_{c}^{d} + \int \left|\frac{d\Phi}{d\tilde r}\right |^2 d\tilde r  + \int V |\Phi|^2 d\tilde r = - \lambda^2 \int |\Phi|^2 d\tilde r .~~~~~
\label{4.7}
\end{eqnarray}
$\tilde r = c $ corresponds to $r=0$ (centre of star).

We see that if Dirichlet or Neumann conditions are imposed at the centre of the star and asymptotically at infinity, then boundary terms go to zero and as $V> 0$, we cannot have any normalizable solutions $\Phi$ satisfying (\ref{4.7}). Note that we have not made any specific assumptions so far on the function $\rho(r)$ which is used to obtain $\mu(r)$. The only detail to be shown is that we can indeed impose Dirichlet or Neumann boundary conditions. For this, let us consider (\ref{4.3}) in the limit $r \to 0$. The analysis of this, and specific solutions need a form of $\rho(r)$. For simplicity, we take $\rho(r) = C$ (constant). Then, $\mu(r) = 4\pi\int_0^r \rho(r)r^2dr = C \frac{4\pi}{3} r^3 $. Then we can approximate (\ref{4.3}) in the limit as $r \to 0$. This reduces to $$\Phi '' -  \frac{l(l+1)}{r^2}\Phi = 0,$$ which has two linearly independent solutions of the form $\Phi = r^{p}$ where $p = \frac{1}{2} \pm \sqrt{l(l+1) + \frac{1}{4}}$. Recalling that $l \geq 2$, Dirichlet boundary conditions are satisfied for the choice $p = \frac{1}{2} + \sqrt{l(l+1) + \frac{1}{4}}$. As $r \to \infty$, it is well-known that we have linearly independent solutions to the Regge-Wheeler equation of the form $e^{\pm \lambda r}$. Thus, choosing the negative sign in the exponent would give us the decaying solution. Therefore we can impose Dirichlet boundary conditions, but there are no normalizable solutions with $\omega = - i \lambda$.\\

For stars, it is more natural to consider quasinormal modes, which are required to be regular at the centre of the star, and outgoing as $r \to \infty$. Astrophysically, we recall that quasinormal modes describe the behaviour of a perturbation of the star which radiates away. This implies that the solution satisfying the quasinormal mode boundary condition must decay in time. Quasinormal modes are in general, complex and the behaviour of the quasinormal mode is one of decaying oscillations. The real part of the quasinormal frequency gives the frequency of oscillation, and the imaginary part, the rate of decay. It needs to be established that any solution to (\ref{4.6}) regular at the centre of the star and outgoing as $ r \to \infty$ has the imaginary part of the right sign so that the solution decays in time . A rigorous proof of this is not easy. We show this in case (ii) for a pure imaginary quasinormal mode.
In case (iii) we will present a heuristic argument for the decay in time for a general complex quasinormal mode. \\

\noindent
\textbf{(ii)} $\omega = - i \lambda $ is a pure imaginary quasinormal mode: \\  The outgoing boundary condition implies for pure imaginary quasinormal modes, that as $r \to \infty$, $\Phi \sim e^{-\lambda r}$. As $r \to 0$, there is only one regular solution of the form $r^{p}$ with $p = \frac{1}{2} + \sqrt{l(l+1) + \frac{1}{4}}$. This is an increasing function. Now, from (\ref{4.6}), replacing $\omega^2 = - \lambda^2$, we see that since $V$ is positive, $\frac{d^2\Phi}{d\tilde{r}^2} > 0$. Therefore, the solution must blow up as $r \to \infty$. We must have $\lambda <0 $. However the time dependence of $\tilde \Omega$ is then of the form $e^{\lambda t}$ indicating that for $\lambda <0 $, the solution decays in time.\\

\noindent
\textbf{(iii)} $\omega = \omega_R + i \omega_{I}$ is a complex quasinormal mode: \\  The outgoing boundary condition implies that for $r$ large, $\Phi \sim e^{-i\omega r} = e^{\omega_I r} e^{-i\omega_{R} r}$. Let us consider (\ref{4.7}) again, where we will now replace $- \lambda^2 $ by $\omega^2 $. Further we take $\rho$ to be constant, which implies that as $r \to 0$, $\tilde r \to 0$. The boundary term coming from integration of parts is not zero, due to the outgoing boundary conditions.
\begin{eqnarray}
\lim_{d \to \infty} i \omega e^{2\omega_{I} d}   + \int_{0}^{d} \left|\frac{d\Phi}{d\tilde r}\right|^2 d\tilde r  + \int_{0}^{d} V |\Phi|^2 d\tilde r = \omega^2 \int_{0}^{d} |\Phi|^2 d\tilde r .~~~~~
\label{4.8}
\end{eqnarray}
The imaginary part of the above equation reads
\begin{equation}
\lim_{d \to \infty} \omega_R \left[ e^{2\omega_{I} d} - 2 \omega_{I}( \int_{0}^{d} |\Phi|^2 d\tilde r ) \right] = 0.
\label{4.9}
\end{equation}
Therefore, if $\omega_R \neq 0$, then we require $\omega_{I} > 0$. From the time dependence of $\tilde \Omega$, this then implies that the perturbation decays in time. The $\omega_R = 0 $ case was already considered before.\\

We provide a note of caution here. This is a heuristic argument, not a rigorous proof. The reason is that we have used the asymptotic `outgoing' form of $\Phi$ while evaluating the boundary term. Also, when taking the limit $d \to \infty$, we see that each of the terms/integrals can blow up. However, the two large terms in (\ref{4.9}) can cancel out each other only if
$\omega_{I} > 0$.

Thus all the above arguments point to the stability of the anisotropic pressure star under axial perturbations. It would also be of astrophysical interest to find the quasinormal mode spectrum for such a star and to see what are the precise signatures in the behaviour of perturbation of the fact that the star has anisotropy in pressure.

\section{Probing instability of a naked singularity}
There are many well-studied examples of matter collapse that ends in a static spacetime describing either a star or a black hole --- a simple example is Oppenheimer-Snyder collapse. Collapse studies of a cloud of anisotropic pressure fluid (with $p_r = 0$) have been shown to approach a variety of anisotropic static spacetimes asymptotically depending on the choice of $p_t $ \cite{joshinarayan}. In particular, one can obtain the class of spacetimes of Florides which were studied in the previous section as the limit of collapse. However, depending on the choice of $p_t$ (or equivalently, choice of equation of state), one can also obtain a naked singularity spacetime. The spacetime is formed as a result of collapse from \emph{regular} initial conditions, and the result is a compact object where the pressure (and curvature) blow up in the centre. The spacetime exterior to the compact object continues to be the Schwarzschild spacetime. The natural question one can ask in view of cosmic
censorship, is whether such pathological spacetimes are obtained in collapse for \emph{generic} initial conditions. This may be difficult to investigate. A simpler problem one can address is the issue of stability of a naked singularity. We expect such pathological spacetimes to be unstable and intuitively expect this to indicate that they do not form from generic initial conditions. In this section, we will discuss the naked singularity spacetime in \cite{joshinarayan} and its stability analysis under axial perturbations.\\

The spherical ball of fluid (of radius $a$) sourcing the naked singularity has density $\rho$ and angular pressure $p_t $ given by
\begin{equation}
 \rho=\frac{M_o}{r^2};~~~p_{t}=\frac{M_o^2}{4r^2(1-M_o)}.
 \label{5.1}
\end{equation}
For the fluid, $p_r=0$. $M_o < 1$ is a parameter related to the total mass of the fluid ball $M_{Total}$ by $M_{o}=2M_{Total}/a$.
For $r\to 0$, both angular pressure and energy density diverge at center showing that $r=0$ should be a curvature singularity. The resulting spacetime consists of an interior metric inside a spherical ball of fluid matched to a Schwarzschild exterior. When $M_o<1$, the curvature singularity is not covered by a horizon, and becomes naked. The spacetime
is then given by
\begin{equation}
ds^{2}=-(1-M_{o})\left(\frac{r}{a}\right)^{\frac{M_{o}}{1-M_{o}}}dt^2+\,\frac{dr^{2}}{1-M_{o}}\,+\,r^{2}d\Omega^{2};
\label{5.2}
\end{equation}
where $r < a$. The exterior metric (for $r > a$)
is given by the Schwarzschild metric written in terms of $M_o$ as
\begin{equation}
ds^2=-\left(1-\frac{M_o a}{r} \right)dt^2+\left(1-\frac{M_o a}{r} \right)^{-1}dr^2+r^2 d\Omega^2 .
\label{5.3}
\end{equation}

In all the discussion that follows, we will have $M_o < 1$ --- for this choice, the fluid satisfies reasonable energy conditions, and yet, we have a naked singularity. Let us now consider axial perturbations of this spacetime. Before we use the analysis of the previous sections, we discuss one important question. Are the equations governing the axial perturbations (or indeed, any wave equations) well-posed in a spacetime that is not globally hyperbolic? The conditions under which the equations would be well-posed in such a case have been discussed by Wald \cite{wald} and others (see \cite{ishiwald}, \cite{horomarolf}, \cite{ishihosoya}, \cite{blau} and references therein). If a mode analysis of the wave equation is possible, then the resulting differential operator needs to be self-adjoint for well-posedness.  In \cite{amrutasuneeta}, the relevant results for proving self-adjointness are given, and as well, a specific naked singularity is shown to be stable under wave perturbations. Thus, we do indeed need to check whether our nakedly singular example is stable or not.

Stability analysis is governed by the master equation (\ref{3.12}). In the collapse study of \cite{joshinarayan}, radial pressure is set to zero and then collapse is studied for various initial conditions. Therefore we do not perturb the vector in the direction of anisotropy. As in the previous section, then, the master equation is homogeneous, and for this static spacetime, on taking $\tilde{\Omega}(r,t)=\Omega(r)e^{\lambda t}$, it is (for $r < a$)
\begin{eqnarray} (1-M_{o})\,\Omega''+\left[\frac{M_{o}}{2r}-\frac{2(1-M_{o})}{r}\right]\Omega'-\left[\frac{\lambda^{2}}{1-M_{o}}\left(\frac{a}{r}\right)^{\frac{M_{o}}{1-M_{o}}}+\frac{m_{V}}{r^{2}}\right] \Omega=0.~~~~~
\label{5.4}
\end{eqnarray}
For $r > a$, it is
\begin{eqnarray}
\left(1-\frac{M_o a}{r}\right)\Omega'' +\left(\frac{3M_o a}{r}-2\right)\frac{\Omega'}{r}-\left[\left(1-\frac{M_o a}{r}\right)^{-1}\lambda^2+\frac{m_V}{r^2}\right]\Omega=0.~~~~~
\label{5.5}
\end{eqnarray}
If we have normalizable solutions $\Omega (r)$ for $\lambda$ real and positive, then $\tilde{\Omega}(r,t)=\Omega(r)e^{\lambda t}$ would grow in time, and we would have an instability. We need to investigate whether there is such an instability.

We can write the interior and exterior equations in Schrodinger form. Let us make the following change of variables: \\
For $r < a$,
\begin{equation}
d\tilde{r}=\frac{r^2}{1-M_o}\left(\frac{a}{r}\right)^{\frac{M_o}{2(1-M_o)}}dr ,
\label{5.6}
\end{equation}
and for $r > a$,
\begin{equation}
d\tilde{r}=r^2\left(1-\frac{M_o a}{r}\right)^{-1}dr .
\label{5.7}
\end{equation}
The Schrodinger equation that results is of the form
 \begin{equation}
 -\frac{d^2\Omega}{d\tilde{r}^2}+V(r)\Omega=0.
 \label{5.8}
\end{equation}
The potential term $V$ for is given by
\begin{eqnarray}
 V(r)&=V_1\quad \quad r< a \nonumber \\& = V_2 \quad \quad r \geq a . \nonumber
\end{eqnarray}
where
\begin{equation}
V_1=\frac{\lambda^2}{r^4}+\frac{m_V(1-M_o)}{r^6}\left(\frac{r}{a}\right)^{\frac{M_o}{1-M_o}} .
\label{5.9}
\end{equation}
\begin{equation}
V_2=\frac{\lambda^2}{r^4}+\frac{m_V}{r^6}\left(1-\frac{M_o a}{r}\right).
\label{5.10}
\end{equation}
These potential can be expressed in terms of $\tilde{r}$ using the transformations given above. Note that for $M_o < 1$, and $\lambda$ real, the potential is positive, and this is a zero eigenvalue problem for a Schrodinger equation with positive potential. By arguments similar to those in the previous section, there are no normalizable solutions to such a problem, provided we can choose Dirichlet or Neumann boundary conditions for $\Omega(r)$ at $r=0$ and asymptotically. Thus, if we can choose one of these boundary conditions, then the naked singularity seems to be stable to axial perturbations.

The behaviour of solutions to (\ref{5.8}) asymptotically as $r \to \infty$ is a linear combination of solutions of the form $r e^{\lambda r}$ and $r e^{-\lambda r}$. To see this, we can just write $\Omega = r\Phi$ and rewrite (\ref{5.5}) for the exterior $r > a$ in Regge-Wheeler form. This asymptotic behaviour shows that there do exist solutions that tend to zero as $r\to \infty$. As $r \to 0$, the behaviour of solutions depends on the value of $M_o$. We give a brief sketch of how solutions depend on $M_o$ and we give the behaviour of the solutions as $r \to 0$ for a range of $M_o$.

In the interior $r < a$, we can integrate (\ref{5.6}) to get $r$ in terms of $\tilde r$. We get
\begin{equation}
r=[k(\tilde{r}+c_1)]^{\frac{2(1-M_o)}{6-7M_o}};\,\, k=\frac{(6-7M_o)}{2}a^{-\frac{M_o}{2(1-M_o)}}.
\label{5.11}
\end{equation}
The integration constant $c_1 $ has to be chosen so that the values of the coordinate $\tilde r$ in the interior and in the exterior agree at $r=a$.

Let $\tilde{r}+c_1=r^*$. The Schrodinger equation for $\Omega$ for  $r< a$ is
\begin{equation}
\frac{d^{2}}{d{r^*}^{2}}(\Omega)-\left[\frac{\lambda^{2}}{(kr^*)^{\frac{8(1-M_o)}{6-7M_o}}}+\frac{m_V(1-M_o)}{a^{\frac{M_{o}}{1-M_{o}}}}\frac{1}{(kr^*)^2}\right]\Omega=0
\label{5.12}
\end{equation}
This equation is of the form
\begin{equation}
-\frac{d^{2}\Omega(r^*)}{d{r^*}^{2}}+\left(\frac{K_{1}}{{r^*}^{f}}+\frac{K_{2}}{{r^*}^{2}}\right)\Omega(r^*)=0;\\
\label{5.13}
\end{equation}
where $f=\frac{8(1-M_{o})}{6-7M_{o}}$, $K_{1}=\frac{\lambda^{2}}{k^{f}}$ and $K_{2}= ( \frac{m_V(1-M_o)}{k^2})a^{\frac{-M_{o}}{1-M_{o}}}$. \\

Let us look at the behavior of this equation for $r \to 0$. We see that $r^* \to 0$ if $6-7M_o>0$ and if $6-7M_o<0$, then $r=0$ is an irregular singular point of (\ref{5.4}). We will leave out this case as it is too complicated to analyze.

\noindent Let $6 - 7M_o > 0$. In the limit of $r^*$ tending to zero, both ${r^*}^{-2}$ or ${r^*}^{-f}$ tend to infinity but one
of them will dominate over the other. We have two possible situations:\\

\noindent
Case I: If $0<M_o \leq \frac{2}{3}$, ${r^*}^{-2}$ is the dominant behaviour in the potential. \\
Case II: If $\frac{2}{3}<M_o<\frac{6}{7}$, ${r^*}^{-f}$ dominates. Again, we have an irregular singular point at $r^{*} = 0$ and we do not consider this case.\\

Let us analyze the case $0<M_o<\frac{2}{3}$.
In this region, ${r^*}^{-2}$ dominates and hence near $r^* \to 0$ boundary our Schrodinger type equation becomes:\\
\[
-\frac{d^{2}\Omega({r^*})}{d{r^*}^{2}}+\frac{K_{2}}{{r^*}^{2}}~\Omega(r^*)=0.
\]\\
Let $\Omega(r^*)={r^*}^{p}$. Plugging this in the above equation, we get two roots of $p$:
\begin{eqnarray}
p_{1}=\frac{1}{2}+\frac{1}{2}(1+4K_{2})^{\frac{1}{2}}>0\\
p_{2}=\frac{1}{2}-\frac{1}{2}(1+4K_{2})^{\frac{1}{2}}<0
\end{eqnarray}
Note that $K_2 > 0$. The general solution is $\Omega(r^*)=C_{1}{r^*}^{p_{1}}+C_{2}{r^*}^{p_{2}}$.
As $r^*\rightarrow0$, ${r^*}^{p_{2}}\rightarrow\infty$ and ${r^*}^{p_1} \to 0$. Imposing Dirichlet boundary conditions, we set $C_{2}=0$. It is also not difficult to verify that with Dirichlet boundary conditions, the Schrodinger operator is self-adjoint.

Thus, we have analyzed the behaviour of axial perturbations in this naked singularity spacetime. For the range $0 < M_o \leq \frac{2}{3}$ we have shown that the spacetime is stable under axial perturbations in a modal analysis, where we have imposed Dirichlet boundary conditions. In fact, for this range of $M_o$, the analysis of axial perturbations is very similar to the stellar interior studied in the previous section. The natural question therefore is: will the polar (scalar) perturbations help distinguish between the well-behaved stellar interior solution of the previous section and this naked singularity? This is a question we hope to investigate in future.

\section{Anisotropic cosmology}
In section III, we considered the most general axial perturbations of the spacetime and anisotropic fluid --- this class included perturbations of the direction of anisotropy. As we saw, perturbations of the anisotropy direction are not physically relevant while dealing with stellar interiors. However, we present an application where such perturbations cannot be neglected. This application is in cosmology, in a situation where the cosmological spacetime is sourced by two or more fluids which may not have a common rest frame. The assumption is that they are noninteracting perfect fluids. It is then possible to show that the energy-momentum tensor of this system can be written as that of a single anisotropic fluid.
Denoting energy density, pressure and velocity of the two fluids by $\rho_{1}, P_{1}, U^{\mu}$ and $\rho_{2}, P_{2}, W^{\mu}$ respectively, the two-fluid energy momentum tensor $$T^{\mu \nu} = (\rho_{1} + P_{1}) U^{\mu}U^{\nu} + P_{1} g^{\mu \nu} + (\rho_{2} + P_{2}) W^{\mu}W^{\nu} + P_{2} g^{\mu \nu}$$ can be brought to the form (\ref{2.2}) by the following steps \cite{Letelier} \footnote{We follow the convention $U^{\mu}U_{\mu} = -1; W^{\mu}W_{\mu} = -1 ;$ which differs from the convention of \cite{Letelier}.}:\\
First we form a pair of vectors $(\bar U^{\mu}, \bar W^{\mu})$ from the pair of four-velocities $(U^{\mu}, W^{\mu})$ such that the expression for $T^{\mu \nu}$ is form-invariant under change from the original pair of four-velocities to the new pair.
\begin{eqnarray}
\bar U^{\mu} &=& (\cos \alpha ) U^{\mu} - \left (\frac{P_2 + \rho_2 }{P_1 + \rho_1 } \right )^{1/2} (\sin \alpha ) W^{\mu};\nonumber \\
\bar W^{\mu} &=& \left (\frac{P_1 + \rho_1 }{P_2 + \rho_2 } \right )^{1/2} ( \sin \alpha ) U^{\mu} -  ( \cos \alpha ) W^{\mu};\nonumber \\
\label{6.1}
\end{eqnarray}
Next, we require that $\bar U^{\mu} \bar W_{\mu} = 0$. This implies that $\alpha $ solves the equation
\begin{eqnarray}
\tan (2 \alpha) = \frac{[(P_1 + \rho_1 )(P_2 + \rho_2 )]}{P_1 + \rho_1 - P_2 - \rho_2 } 2 W^{\mu}U_{\mu}.
\label{6.2}
\end{eqnarray}
It can be checked that $\bar U^{\mu}$ is a timelike vector, and therefore $\bar W^{\mu}$ is spacelike. Lastly, defining the following:
\begin{eqnarray}
V^{\mu} &=& \bar U^{\mu}/(- \bar U^{\nu} \bar U_{\nu} )^{1/2} ;\nonumber \\
X^{\mu} &=& \bar W^{\mu}/( \bar W^{\nu} \bar W_{\nu} )^{1/2} ;\nonumber \\
\rho &=& T_{\mu \nu} V^{\mu} V^{\nu} ; \nonumber \\
P_r &=& T_{\mu \nu} X^{\mu} X^{\nu} ; \nonumber \\
P_t &=& P_1 + P_2 ;
\label{6.3}
\end{eqnarray}
we can check that $T^{\mu}_{\nu}$ for the two-fluid system can be written in the form (\ref{2.2}). Thus the vector $V^{\mu}$ is the velocity of this anisotropic fluid, and if $X^{\mu}$ (the vector in the direction of anisotropy) is in the radial direction, then $P_r $ is the radial pressure and $P_t $ the angular pressure.

The purpose of writing these expressions is to see that if we now perturb the velocities of the two-fluid system, this perturbs both $V^{\mu}$ and $X^{\mu}$. Thus we have to consider both while studying the axial perturbations of the anisotropic fluid.

Let us now look at examples of resulting cosmological spacetimes which have spherical symmetry with respect to the frame in which the anisotropic fluid is at rest (as can be checked, in this frame, both the actual cosmological fluids are moving).
Spacetime metrics have been found which reduce to one of the Friedmann models in some limit (large $r$ or large $t$) \cite{bayin} and have the form:
\begin{equation}
ds^2 = - dt^2 + q(t)[f(r) dr^2 + r^2 d\Omega^2].
\label{6.4}
\end{equation}
Here $q(t)$ is a function which solves the equation $$q \ddot q -\frac{1}{4} {\dot q}^{2} = c q$$ where $c$ is an arbitrary constant. Different choices of $f(r)$ correspond to choices of energy density and pressures.
If we now consider axial perturbations of this class of spacetimes as outlined in section III, then the equation for conservation of the perturbed energy-momentum tensor (\ref{3.10}) becomes
\begin{eqnarray}
q(t) r [\dot \tau^{0} + \tau^{1'} ] + 2 r \dot q \tau^{0}  +
\frac{ rf'}{2f} q \tau^{1} + 3 q \tau^{1} = 0.
\label{6.5}
\end{eqnarray}
In terms of the perturbations of the vectors $\delta V_{i}=\tilde{\alpha}\mathcal{V}_{i}$ and $\delta X_{i}=\tilde{\beta}\mathcal{V}_{i} $, $\sqrt{q} r \tau^{0} = (\rho + p_t ) \tilde \alpha$ and $\sqrt{q} r \tau^{1} = (p_r - p_t ) \frac{1}{\sqrt{mf}} \tilde \beta $. In this problem, we will have to consider both $\tilde \alpha, ~\tilde \beta \neq 0$. This corresponds to perturbing the velocities of both fluids in the original two-fluid system. Even if we perturb only one of the two original cosmological fluids, this implies both $\tilde \alpha, ~\tilde \beta \neq 0$. The perturbations can be expressed in terms of the master equation (\ref{3.12}) --- which will now be nonhomogeneous. The right-hand side is not zero as was the case in examples from previous sections. Our purpose in this section was not to analyze stability of any specific anisotropic cosmologies --- this would be complicated both due to the nonhomogeneous term and the fact that the background metric is not static (so a modal stability analysis may not be possible). It was rather to illustrate an example where perturbations of the `direction of anisotropy' are relevant.

\section{Summary}
In this paper, we have begun a systematic study of perturbations of spacetimes sourced by fluids with pressure anisotropy. The class of spacetimes we considered have spherical symmetry, which allows breaking the problem into axial and polar perturbations. We have discussed the axial case in this paper and work on the polar perturbations is in progress. As we have shown, the most general class of axial perturbations involves perturbing the direction of anisotropy. In such a scenario, axial perturbations can excite/be excited by fluid perturbations. This is in contrast to the perfect fluid case. The resulting equations for perturbations can all be simplified to one inhomogeneous `master' equation for a gauge invariant perturbation variable (a function) using the formalism developed in \cite{ishikodama}. If the direction of anisotropy is not perturbed, then, as in the case of perfect fluids, axial perturbations cannot excite fluid perturbations. The master equation in such a case is homogeneous, and if the spacetime is static, can be reduced to a Schrodinger-type ODE. For stellar interior spacetimes, it is natural to keep the direction of anisotropy constant. However, there are other examples such as anisotropic cosmologies, where we have shown that we must consider perturbations of the direction of anisotropy as well. We have discussed the master equation for a class of anisotropic cosmologies.

As an application of our results, we have also considered two spacetimes, which can be obtained as the endpoint of collapse of a cloud of anisotropic fluid. The first spacetime is nonsingular, and can model a stellar interior. The second is pathological and has a naked singularity. For the stellar interior, we prove its stability under axial perturbations both for Dirichlet and quasinormal mode boundary conditions on the perturbation. In principle, our equations could also be used to numerically obtain the quasinormal mode spectrum of the star with this interior. This could provide astrophysical signatures of the anisotropy in the interior. The analysis of axial perturbations of the naked singularity shows striking resemblances to the previous well-behaved stellar interior. In particular, the singular spacetime (for a range of mass) is also stable under axial perturbations. This intriguing similarity is a motivation for studying stability under polar perturbations. Our expectation is that the singular spacetime should be unstable, while the stellar interior ought to be stable --- it would be interesting to see if the polar perturbations reveal some surprises. In fact, there is already a hint from earlier work of Chan, Herrera and Santos \cite{chan2}, \cite{chan3} (which includes shear and viscous effects) and \cite{chan1} of instabilities under a sub-class of polar perturbations \footnote{ We thank the referee for bringing these papers to our attention.}. These authors have analyzed stability of stellar interiors under a sub-class of polar perturbations where even the temporal behaviour is derived from the linearized perturbation equations. They find a link between the pressure anisotropy of the background spacetime and (in)stability independent of choice of equation of state. Thus stability analysis under general polar perturbations promises to be much more complex. The natural question is whether there exist bounded unstable polar perturbations for various values of initial pressure anisotropy that lead to collapse. We are working on the polar perturbation case and hope to report on some of these issues in the near future.

The computations in this paper could also be done for higher dimensional spacetimes. In that case, we would have to consider additional tensor perturbations on the $n$-sphere part of the metric. However, the most important generalization of this work would be to anisotropic spacetimes which do not have spherical symmetry. Spacetimes with pressure anisotropy along one cartesian direction and isotropic pressure in the plane tangent to this direction have interesting physics applications. They appear as duals to anisotropic RHIC plasma \cite{janik}, \cite{mateos} and in cosmology \cite{harko}. A systematic study of perturbations of such spacetimes needs to be done to see if the full stability problem can be broken into classes of perturbations which decouple from each other.


\begin{thebibliography}{100}
\bibitem{herrera} L. Herrera, N. Santos, Phys. Reports 286 \textbf{(1997)} 53.
\bibitem{florides} P.S. Florides, Proc. Roy. Soc. London Ser.A, 337 \textbf{(1974)} 247.
\bibitem{gleiser} K. Dev, M. Gleiser, Gen. Rel. Grav. 35 \textbf{(2003)} 1435.
\bibitem{joshinarayan} P.S. Joshi, D. Malafarina, R. Narayan, Class. Quant. Grav. 28 \textbf{(2011)} 235018.
\bibitem{kyriakopoulos} E. Kyriakopoulos, Int. J. Mod. Phys. D22 \textbf{(2013)}.
\bibitem{mazur} P. Mazur, E. Mottola, Proc. Nat. Acad. Sci. 101 \textbf{(2004)} 9545.
\bibitem{pedraza} P.H. Nguyen, J.F. Pedraza, Phys. Rev. D88 \textbf{(2013)} 064020.
\bibitem{Letelier} P.S. Letelier, Phys. Rev. D22 \textbf{(1980)} 807.
\bibitem{Letelier1} P.S. Letelier and P.S.C. Alencar, Phys. Rev. D34 \textbf{(1986)} 343.
\bibitem{bayin} S.S. Bayin, Ap. J. 303 \textbf{(1986)} 101.
\bibitem{harko} T. Harko, F.S.N. Lobo, JCAP 07 \textbf{(2013)} 036.
\bibitem{solana} J. Casalderrey-Solana, H. Liu, D. Mateos, K. Rajagopal, U.A. Wiedemann, \emph{Gauge/String Duality, Hot QCD and Heavy Ion Collisions}, Cambridge University Press, \textbf{(2014)}.
\bibitem{janik} R.A. Janik, P. Witaszczyk, JHEP0809 \textbf{(2008)} 028.
\bibitem{mateos} D. Mateos, D. Trancanelli, Phys. Rev. Lett. 107 \textbf{(2011) }101601.
\bibitem{ruggeri} L. Herrera, G.J. Ruggeri, L. Witten, Ap. J. 234 \textbf{(1979)} 1094.
\bibitem{chan2} R. Chan, L. Herrera, N. O. Santos, Mon. Not. Roy. Astr. Soc. 265 \textbf{(1993)} 533.
\bibitem{chan3} R. Chan, L. Herrera, N. O. Santos, Mon. Not. Roy. Astr. Soc. 267 \textbf{(1994)} 637.
\bibitem{chan1} R. Chan, L. Herrera, N. O. Santos, Class. Quant. Grav. 9 (1992) L133.
\bibitem{gleiser2} K. Dev, M. Gleiser, Gen. Rel. Grav. 35 \textbf{(2003)} 1435.
\bibitem{herreragrg} L. Herrera, N.O. Santos, Gen. Rel. Grav. 27 \textbf{(1995)} 1071.
\bibitem{herrerapla} L. Herrera, A. Di Prisco, J.L. Hernandez-Pastora, N.O. Santos, Phys. Lett. A 237 \textbf{(1998)} 113.
\bibitem{tolman} R. Tolman, Phys. Rev. 35 \textbf{(1930)} 875.
\bibitem{whittaker} E.T. Whittaker, Proc. Roy. Soc. Lond. A149 \textbf{(1935)} 384.
\bibitem{chirenti} C. Chirenti, L. Rezzolla, Class. Quant. Grav. 24 \textbf{(2007)} 4191.
\bibitem{cardoso} V. Cardoso, P. Pani, M. Cadoni, M. Cavaglia, Phys. Rev. D77 \textbf{(2008)} 124044.
\bibitem{chirenti2} C. Chirenti, L. Rezzolla, Phys. Rev. D78 \textbf{(2008)} 084011.
\bibitem{perko} T.E. Perko, R.A. Matzner, L.C. Shepley, Phys. Rev. D6 \textbf{(1972)} 969.
\bibitem{den} K. Tomita, M. Den, Phys. Rev. D34 \textbf{(1986) }3570.
\bibitem{miedema} P.G. Miedema, W.A. van Leeuwen, Phys. Rev. D47 \textbf{(1993)} 3151.
\bibitem{mateos2} D. Mateos, D. Trancanelli, JHEP1107 \textbf{(2011)} 054.
\bibitem{cheng} L. Cheng, X-H. Ge, S-J. Sin, Phys. Lett. B734 \textbf{(2014)} 116.
\bibitem{cheng1} L. Cheng, X-H. Ge, S-J. Sin, \emph{Anisotropic plasma at finite U(1) chemical potential},  arXiv: 1404.5027.
\bibitem{ishikodama} A. Ishibashi, H. Kodama, Prog. Theor. Phys. Suppl. 189 \textbf{(2011)} 165.
\bibitem{misner} C.W. Misner, D.H. Sharp, Phys. Rev. 136 \textbf{(1964)} B571.
\bibitem{chandraferrari} S. Chandrasekhar, V. Ferrari, Proc. Roy. Soc. London Ser.A, 432 \textbf{(1991)} 247.
\bibitem{herreraPRD} L. Herrera, J. Ospino, A. Di Prisco, Phys. Rev. D77 (2008) 027502.
\bibitem{bowersliang} R. Bowers, E.P. Liang, Ap. J. 188 \textbf{(1974)} 657.
\bibitem{sawyer} R.F. Sawyer, D.J. Scalapino, Phys. Rev. D7 \textbf{(1973)} 953.
\bibitem{reggewheeler} T. Regge, J.A. Wheeler, Phys. Rev. 108 \textbf{(1957)} 1063.
\bibitem{wald} R.M. Wald, J. Math. Phys. 21 \textbf{(1980)} 2802.
\bibitem{ishiwald} A. Ishibashi, R.M. Wald, Class. Quant. Grav. 20 \textbf{(2003)} 3815.
\bibitem{horomarolf} G.T. Horowitz, D. Marolf, Phys. Rev. D52 \textbf{(1995)} 5670.
\bibitem{ishihosoya} A. Ishibashi, A. Hosoya, Phys. Rev. D60 \textbf{(1999)} 104028.
\bibitem{blau} M. Blau, D. Frank, S. Weiss, JHEP 0608:011 \textbf{(2006)}.
\bibitem{amrutasuneeta} A. Sadhu, V. Suneeta, IJMPD 22 (2013) 1350015.
\end{thebibliography}
\end{document}